\documentclass[aps,prl,twocolumn,superscriptaddress,nofootinbib,nobalancelastpage,showpacs,nobibnotes]{revtex4}
\usepackage{epsfig}

\begin{document}
\title
{Speed-up of neutrino transformations in a supernova environment.
}

\author{R. F. Sawyer}
\affiliation{Department of Physics, University of California at
Santa Barbara, Santa Barbara, California 93106}

\begin{abstract}
When the neutral current neutrino-neutrino interaction is treated completely, rather
than as an interaction among angle-averaged distributions, or as a 
set of flavor-diagonal effective potentials, the result can be flavor
mixing at a speed orders of magnitude faster than that one would anticipate from
the measured neutrino oscillation parameters.
It is possible that the energy spectra
of the three active species of neutrinos emerging from a supernova are
nearly identical. 


\end{abstract}
\maketitle

\section{Introduction}

Predictions regarding the neutrino flavor physics in the region of the supernova's neutrino-sphere have remained quite stable as treatments of other issues have evolved over time. These predictions, which involve significantly different
energy spectra for the $\nu_e$, $\bar \nu_e$, $\nu_{\mu,\tau}$, $\bar \nu_{\mu, \tau}$,  are important to the theory of
the region above the neutrino-sphere. They matter for the explosion dynamics, for heavy element
nucleosynthesis, and for analyzing the neutrino burst from the galactic supernova
scheduled for some time in the twenty-second century. 

As the three known species of neutrinos move through layers of the supernova
that are immediately under the effective neutrino-sphere, their 
physics can be much more complex than is assumed
in current simulations of the whole supernova process. Before describing some
of these complexities in detail, we look at parameters that describe the region:
taking a density of $10^{11} {\rm g c^{-3}}$ and temperature of $7$ MeV.
There is a characteristic energy that we define from the Fermi
constant and the neutrino number density, $E_s \equiv 2 \sqrt 2 G_F n_\nu $, where $n_\nu$ is the neutrino number density.
which when translated into distance, under the above conditions, gives 
$E_s^{-1}=.04$cm \footnote{We use units in which $\hbar=1$, $c=1$} . Of course, this is many orders of magnitude less than the distance set by the density and
cross-section (proportional to $G_F^{-2}$) , a distance scale that is tens of km., the scale of 
the neutrino-sphere radius itself.

However, terms of order $E_s$ do enter in forward (or ``index of refraction") effects. In a 
medium containing electrons, the $\nu$-$e$ interaction energies create different potentials for $\nu_e$'s, and for $\bar \nu_e$'s, 
than for $\nu_\mu$'s and $\nu_\tau$'s and their antiparticles. The induced terms of order $E_s$ enter interesting physics in conjunction with the 
($\delta m^2$) neutrino oscillation parameters,
but these oscillation effects do not enter significantly out to a distance of many times the neutrino-sphere radius, if we use current
values for the neutrino mass parameters and operate in the commonly accepted theoretical framework. The electron densities 
have no effect on the  oscillation between $\mu_\nu$ and $\mu_\tau$, with oscillation 
length of the order of a kilometer for a $20$ MeV neutrino, and uninteresting in any case by virtue of the near identities of their spectra snd angular distributions. The $\nu_e$ oscillation length in vacuum is 
longer by a factor of 40 or so; furthermore, the large electron density ensures that oscillation amplitudes will
be miniscule. Thus there is negligible ``forward" neutrino physics of any consequence
in the conventional picture of this region, barring a sterile neutrino that is attached to a much 
bigger mixing parameter.

All of this can change when we include the full neutral current interaction among the neutrinos, however. The
effective forward interaction Hamiltonian for the system now contains terms that can exchange
neutrino flavors. At first sight this seems fairly irrelevant in the region below the 
neutrino-sphere. The initial system coming from
the supernova core can be taken as an incoherent sum of flavor eigenstates. Then the  mixing of the $\nu_e$ remains small, 
according to the usual angle-averaged equations, which assume isotropic momentum distributions.
A large, nearly flavor diagonal, surrounding neutrino density can serve to synchronize \cite{s1}-\cite{s8} these
small oscillations, according to these equations, but not to enhance their amplitude when the amplitudes otherwise would be small.  

In a region well above the neutrino-sphere the oscillation parameters can produce
strong mixing and possible MSW transformations.
Considerable attention has been given to the effects of the neutralcurrent coupling terms, including
the forward flavor-exchange parts, in this region \cite{sig}-\cite{qf}. Here neutrino anisotropy was taken into account by using the isotropic form for the equations, but
with an effective coupling constant that
incorporates a single (``flux-averaged") angle of incidence for $\nu$-$\nu$ encounters.
However we now know that other non-linear phenomena can take over when
anisotropy is incorporated into the evolution equations in a more correct way \cite{rfs1}. These phenomena, to some degree, 
can see the time scale $t_s= E_s^{-1}$ and overturn our intuitions about the problem.

In the present paper we look at the neutrino physics only over short distances, over which
the surrounding density and temperature can be taken to be constant.
Even this abstracted problem is extraordinarily complex. 
The simulations in \cite{rfs1}, backed up by analytic considerations, were based on 
two-flavor models with simple anisotropies, essentially a few clusters of definite direction in momentum space.
They show the possibility of one kind of instability, in the form of an exponentially increasing
growth in flavor-mixing amplitudes seeded by the tiniest (coherent) initial mixing (which could come from
earlier neutrino oscillation effects). In these examples the rise-time of the exponential is the shortest time scale
in the problem, of order $E_s^{-1}$.
This behavior is in contrast to the oscillatory behavior in the isotropic
approximation. That said, in the domains considered in ref.\cite{rfs1} the conditions that can lead to this runaway evolution are delicately
dependent on the initial flavor-momentum correlations, and are subject to change when
new elements are added to the model.

Approaching an authoritative answer in a more realistic supernova environment
 requires the inclusion of all three active flavors and their antiparticles.\footnote{Inclusion of a sterile species would
complicate our considerations enormously.}
Let us think of the neutrino distributions in the transitional
region just below the average depth of the last $\nu_\mu$ and $\nu_\tau$ scatterings, a region that is
somewhat farther below the average depth of the last $\nu_e$ scattering. In this region the 
$\nu_e$ momentum distribution is fairly isotropic, whereas the $\nu_\mu$,$\nu_\tau$, $\bar \nu_\mu$, $\bar \nu_\tau$
distributions are already quite biased toward upward momenta. Both in angular distribution and 
in spectrum, the $\bar \nu_e$'s lie between the $\nu_e$'s and the $\nu_{\mu,\tau}$'s. In our first idealization we picture this mixture
as a core group of all six species in equal number with a momentum
distribution peaked somewhat upwards, superimposed on a group in which we have $\nu_{\nu, \tau}$'s and their
anti-particles in a 
distribution peaked strongly upward, and $\nu_e$, $\bar \nu_e$'s in a distribution peaked downward.
We call the second group the valence group. The core group is irrelevant to the flavor evolution
of the system, as may be obvious but in any case is supported by our later formalism.
The problem that we now solve is one in which we replace the valence system by narrow angle
beams of $\nu_\mu$ and $\nu_\tau$ neutrinos and antineutrinos directed outward, and beams of $\nu_e$'s and 
$\bar\nu_e$'s directed inwards. In a second version we shall assign the $\bar \nu_e$'s to the outgoing group, corresponding 
to a less outwardly biased core group.

Over the time-scales that we consider below, the interactions that come into play, in addition to the
neutral current couplings, are the neutrino mass-mixing terms and the difference between the forward e-$\nu_e$ and e-$\nu_{\mu,\tau}$
scattering amplitudes.
These interactions would not even matter over our time scales, but for the instabilities
in the non-linear equations that govern the evolution of the system. The first initial state that we pick consists of down-moving
$\nu_e$'s and $\bar \nu_e$'s, each with a number density of $n_v$, with the other four
species of particles moving upward each with the same number density. The outcome, in this case and
in two other cases with somewhat different initial conditions, is a
scenario of extremely rapid mixing of flavors, leading to a rapid destruction of some of the initial correlations of flavor
with angle, and in turn with energy spectrum.

\section{Equations of evolution}

The equations are an extension of the equations of Sigl and Raffelt \cite{sr}. The main new element is the explicit inclusion
of all three flavors of active neutrinos and antineutrinos, as well as perhaps a more  incisive inclusion of angular factors,
so that we can better treat anisotropy effects. The formalism is much simpler than that presented in ref.\cite{sr} because
of our limitation to purely forward effects, justified by the short time scales to be considered. 

All of the dynamics of the system can be described in terms of the annihilation
and creation operators for the three neutrino flavors and the momentum states of the initial
configuration; we define the annihilators of the mode $p$ for the respective three flavors $\nu_e,\,\nu_\tau,\,\nu_\mu$ as $a_p,b_p,c_p$
with the annihilators for the antiparticles denoted $\bar a_p,\bar b_p,\bar c_p$.
The neutral-current effective interaction Hamiltonian, to be exhibited below, is independent of
the neutrino energies and depends only on relative directions of motion. Therefore, for the purpose of discussing 
the evolution of the density matrix due to this interaction alone, we can bundle all energies 
together and define collective operators for each angle. We divide the solid angle of the momentum space into
elements $d \Omega$ centered at angle $\Omega$ and define collective operators for each of these elements
using a shorthand symbol $S_{\Omega}$ which operates according to the rule,

\begin{equation}
S_{\Omega}[a^\dagger b]=(d\, \Omega)^{-1} \sum_{p \,\subset \,{d \Omega }}a_p^\dagger b_p \, .
\end{equation}
We choose,
\begin{eqnarray}
\rho_1(\Omega )   =S_\Omega [a^\dagger b ]~~,~~\rho_2(\Omega)
=S_\Omega [b^\dagger a]~,
\nonumber\\
\,
\nonumber\\
\rho_3(\Omega)=S_\Omega [(2 c^\dagger c -a^\dagger a -b^\dagger b)/3]~,
\nonumber\\
\,
\nonumber\\
 ~\rho_4(\Omega )=S_\Omega [b^\dagger c]~~, ~~ \rho_5(\Omega)=S_\Omega [c^\dagger b]~,
\nonumber\\
\,
\nonumber\\
\rho_6 (\Omega )=S_\Omega [(2 a^\dagger a -b^\dagger b -c^\dagger c)/3]\, ,
\nonumber\\
\,
\nonumber\\
 \rho_7(\Omega ) =S_\Omega [c^\dagger a]~, ~ \rho_8 (\Omega)
=S_\Omega [a^\dagger c]~,
\nonumber\\
\,
\nonumber\\
\rho_9(\Omega )=S_\Omega [(2 b^\dagger b-a^\dagger a -c^\dagger c)/3]~ ,
\nonumber\\
\,
\nonumber\\
\rho_0 (\Omega)=S_\Omega [ a^\dagger a+ b^\dagger b+c^\dagger c]\,.
\label{ops}
 \end{eqnarray}
Given the redundancy $\rho_3+\rho_6+\rho_9=0$, these embody the usual SU3 relation, $3^* \times 3=1+8$.
We construct bilinears of the antiparticle operators $\bar a,\bar b,\bar c$ in the same fashion as above except that we  make the exchanges in indices, relative to the neutrino case, of $1\leftrightarrow2$, $4\leftrightarrow 5$, and $7 \leftrightarrow 8$. For example, we take,
\begin{equation}
\bar \rho_1 (\Omega)=S_\Omega [\bar b^\dagger\, \bar a ]~~
,~~\bar \rho_2 (\Omega)=S_\Omega [ \bar a^\dagger  \, \bar b] ~,~{\rm etc.} 
\end{equation}
The commutation rules are now given by,
\begin{eqnarray}
[\rho_i (\Omega), \rho_j (\Omega ')]=\delta (\Omega-\Omega ') \sum_{k=1}^9 f_{i,j,k} \rho_k (\Omega)\, ,
\nonumber\\
\,
\nonumber\\
\, [ \bar \rho_i (\Omega),\bar \rho_j (\Omega ')]=-\delta (\Omega-\Omega ') \sum_{k=1}^9 f_{i,j,k}\bar \rho_k (\Omega) \, ,
\nonumber\\
\,
\nonumber\\
\, [\bar \rho_i (\Omega), \rho_j (\Omega ')]=0 ~.~~~~~~~~~~~~
\label{commutators}
\end{eqnarray}

The $f_{i,j,k}$ coefficients 
characterize the SU3 algebra of the operator arguments of the expressions (\ref{ops}) as they describe a single
momentum state. They are listed in the appendix.

In terms of these operators the forward neutral current interaction Hamiltonian is, \footnote{
The kinetic term in the Hamiltonian takes the same value for every state in our set,
and therefore can be ignored in all that follows.}
\begin{eqnarray}
&H_{\rm for}={ G_F \over {\sqrt 2 \rm (Vol.)}}\int d\Omega  \, d\Omega'(1-\cos \theta_{\Omega,\Omega'})
\nonumber\\
&\bigr \{{4 \over 3} [\rho_0 (\Omega)- \bar \rho_0 (\Omega)]\times  [\rho_0 (\Omega ')- \bar \rho_0 (\Omega ')]+
\nonumber\\
& \sum_{i=1}^9 [ \rho_i (\Omega )- \bar \rho_i (\Omega )]  [\rho_i^\dagger( \Omega ') -\bar \rho_i^\dagger (\Omega ')] 
\bigr \}\, .
\label{ham}
\end{eqnarray}
The result (\ref{ham}) is obtained from the full neutral current four-Fermion interaction 
by retaining only those combinations of operators that give a forward amplitude, ${\bf p,q}\rightarrow
{\bf p,q}$ , and restricting these momenta to the set of states that are initially occupied. 
The terms in (\ref{ham}) for
$\nu$-$\nu$ (and $\bar \nu$-$\bar \nu$) scattering come from two graphs. In one of these a zero-momentum $Z$ 
is exchanged, giving rise to an interaction among the $\rho_0$ and $\bar \rho_0$ operators only, since for this term
in the scattering ${\bf p+q} \rightarrow {\bf p+q}$, the operator that makes the final ${\bf p}$ and the operator that annihilates the
initial ${\bf p}$ attach to the same end of the $Z$ line. This singlet-singlet coupling (in the SU3 sense) gives
 3/4 of the first term in (\ref{ham}) and contributes nothing to the time evolution of the density matrix.
In the other  (``crossed") term for $\nu$-$\nu$ scattering, a momentum $\bf {p-q}$ is transferred through the $Z$.
In this term we use a Fierz identity, giving a multiplying minus sign,
one anticommutation, giving a second minus sign, and finally we use the appropriate crossing matrix in the internal
SU3 space to produce the rest of the terms in (\ref{ham}). For the particle-anti-particle terms, the crossing is
slightly different, since the ``crossed" graph is a virtual annihilation into an intermediate $Z$, in which the $Z$ carries momentum $\bf {p+q}$. Our $1\leftrightarrow 2$, etc., exchange in subscripts in defining the anti-particle 
densities reflects this change. Of course, since the neutrinos will be in the MeV energy range, we can neglect
the neutrino momenta in the $Z$ propagator.

To the above Hamiltonian we add
a neutrino mass term, taking the $\nu$ and the $\bar \nu$'s all to have 
nearly enough a common energy, $E$, so that the oscillation term is the same for each. (We
weaken this assumption later.) We choose,
\begin{eqnarray}
&H_{\rm mass}= \lambda_1 \int  d\Omega [ \rho_1 (\Omega)+\rho_2 (\Omega) +\bar \rho_1(\Omega)+\bar \rho_2 (\Omega)]+
\nonumber\\
 &\lambda_2 \int d \Omega  [ \rho_4 (\Omega)+\rho_5 (\Omega) +\bar \rho_4(\Omega)+\bar \rho_5 (\Omega)]\, ,
\label{osc1}
\end{eqnarray} 
where $\lambda_1$ is given by $(2 E)^{-1}$ times the $\delta m^2$ parameter for $\nu_{e,\tau}$ mixing and
$\lambda_2$ is given by $(2 E)^{-1}$ times a $\delta m^2$ parameter for $\nu_{\mu,\tau}$ mixing.
Finally we include the (relative) energies of interaction
that result from a net $ n_{e}-\bar n_{e}$ electron-positron density difference,
\begin{equation}
H_e=\sqrt 2 G_F (n_{e}-\bar n_{e}) \int d \Omega[ \rho_6 (\Omega) - \bar \rho_6 (\Omega)]\, .
\label{osc2}
\end{equation}

From the above Hamiltonian and commutation rules we can write Heisenberg equations for our set of operators $\rho_i (\Omega)$, $\bar \rho_i (\Omega)$ . The right hand sides will be quadratic functions of the densities themselves,
with one angular integration. Thus \underline{if} we are allowed to replace the expectation values
of products of density operators, on the right hand sides, by products of expectation values, we obtain a closed set
of non-linear, integro-differential equations for the expectations. More explicitly, we take an initial condition
in which the expectation of any product of  our density operators is the
product of the expectations of the individual operators, e. g.,
\begin{equation}
 \langle  \rho_3 (\Omega),t) \rho_6(\Omega ,t) \rangle=
\langle \rho_3(\Omega ,t)  \rangle \langle \rho_6(\Omega, t)  \rangle \, ,
\end{equation}
at $t=0$. The further assumption is that this factorization holds
for all subsequent times. In the present work we assume that this is justified.\footnote{Truncation schemes such as this should be regarded with some suspicion although we note that the whole literature pertaining to the effects of $\nu-\nu$ interactions is based on the factorization assumption. In previous work \cite{rfs1}, \cite{rfs2}, in cases
of simpler algebras, we have found that the defect of the expectation-value assumption can be expressed
as additional terms proportional to $N_\nu^{-1}$in non-linear equations for the \underline{
products} of density operators. However we also found
that when the non-linear background equations had a certain instability, these terms could lead to
rates of otherwise forbidden processes that are suppressed only by a factor $\log (N)$, and may need
to be retained even when $N$ is huge.} To illustrate the complexity of our system, we write the first of the ten independent equations
of evolution, where now all density variables are understood as expectation values,

\begin{eqnarray}
&i\dot \rho_1(\Omega)={G_F \over \sqrt2 {\rm Vol.}}\int d \Omega ' \, (1-\cos \theta_{\Omega , \Omega ' })\Bigr \{\rho_7^*(\Omega )[\rho_4^*(\Omega ')-
\nonumber\\
&\bar \rho_4^*(\Omega ')]+\rho_1(\Omega) [\rho_9(\Omega ')- \rho_6 (\Omega') -\bar \rho_9(\Omega' )+\bar \rho_6 (\Omega ')]-
\nonumber\\
&\rho_4 (\Omega)^*
[\rho_7^* (\Omega')-\bar \rho_7^*(\Omega ')]+[\rho_6 (\Omega)-\rho_9 (\Omega )][\rho_1 (\Omega ')-
\nonumber\\
&\bar \rho_1 (\Omega')] \Bigr \}+\lambda_1 [\rho_6 (\Omega)-\rho_9 (\Omega) ]+\lambda_2 \rho_7^*(\Omega) -
\nonumber\\
&{\sqrt 2 G_F (n_e -\bar n_e) }\rho_1 (\Omega)\, .
\nonumber\\
\label{hummer}
\end{eqnarray}

The equations are unmanageable except in simplified
situations. As mentioned above, there have been a number of treatments of the two-flavor case in which the angular 
distributions are isotropic. In work reported below
we have found non-linear phenomena for the three-flavor case that are qualitatively different from the 
``synchronized" oscillations that can arise in the two-flavor system, even in the isotropic case. We return to this point, 
but we include 
anisotropy in all of the detailed solutions that we present later. Now we implement the two-beam idealization, as discussed in the
introduction. Setting $1+\cos \theta_{\Omega, \Omega '}=2$, we define collective density 
coordinates for the two beams as, 

\begin{eqnarray}
x_i=(N_\nu)^{-1}\int_{\rm down}d \Omega\, \rho_i (\Omega)\, ,
\nonumber\\
y_i=(N_\nu)^{-1}\int_{\rm up} d \Omega \, \rho_i (\Omega)\, ,
\end{eqnarray}
where $N_\nu=n_\nu ({\rm Vol.})$ is the initial number of each flavor of $\nu$, in our ``valence" group,
and $n_\nu$ is the corresponding number density. We define $\bar x_i$, $\bar y_i$ in parallel
fashion.
The commutation rules are now of the form,
\begin{eqnarray}
[x_i, x_j]=(N_\nu)^{-1} \sum_k f_{i,j,k} x_k~ ,
\nonumber\\
\,
\nonumber\\
\,[\bar x_i,\bar x_j]=-(N_\nu)^{-1} \sum_k f_{i,j,k} \bar x_k ~,
\label{com2}
\end{eqnarray}
The $y_i, \bar y_i$ variables have the same commutation rules among themselves, and 
$[x_i, y_j]=0$.

The operative neutral-current Hamiltonian for these variables is,

\begin{eqnarray}
\,
\nonumber\\
H_{\rm for}={2 \sqrt 2 N_\nu^2 G_F \over { \rm (Vol.)}}
\bigr \{{4 \over 3} [x_0 - \bar x_0 ]
[y_0 - \bar y_0 ]
\nonumber\\
+ \sum_i^9 [ x_i - \bar x_i ]  [y_i^\dagger -\bar y_i^\dagger] \bigr \}~.~~~~~~~~~~~
\label{hama}
\end{eqnarray}
The $\nu$-mass and electron interaction terms are given by,
\begin{eqnarray}
&H_{\rm mass}+H_{e}= N_\nu 2 \sqrt 2 G_F n_\nu [ \lambda_1' (x_1+x_2+y_1+y_2+
\nonumber\\
&\bar x_1+\bar x_2+\bar y_1 +\bar y_2)+\lambda_2' (x_4+x_5+y_4+y_5+\bar x_4 +
\nonumber\\
 &\bar x_5+\bar y_4+\bar y_5)+
{(n_e-\bar n_e)\over 2 n_\nu}( x_6+y_6-\bar x_6 -\bar y_6)]~,~
\label{hamb}
\end{eqnarray}
where dimensionless oscillation parameters have been introduced, $\lambda_i'=\lambda_i/(2\sqrt 2 G_F n_\nu)$.
Writing the Heisenberg equations that come from the total Hamiltonian $H_{\rm for}+H_{\rm mass}+H_e$,
and the commutation rules (\ref{com2})and measuring time in units $(2 \sqrt 2 G_F n_\nu )^{-1}$, we obtain a set of 24 coupled 
nonlinear equations for the 24 density matrix components. We show the first two here; more are provided in the appendix,
\begin{eqnarray}
&i\dot x_1=x_1(y_9-y_6-\bar y_9+\bar y_6)-x_4^*(y_7^*-\bar y_7^*)+
\nonumber\\
&x_7^*(y_4^*-\bar y_4^*)+(x_6-x_9)(y_1-\bar y_1)+\lambda_1' (x_6-x_9)+
\nonumber\\
&\lambda_2' x_7^* -{n_e -\bar n_e \over 2 n_v}x_1 \, ,
\nonumber\\
\,
\nonumber\\
&i \dot x_3=x_4^*(y_4-\bar y_4)-x_4(y_4^*-\bar y_4^*)+~~~~~~~~~~~~~~~~
\nonumber\\
&x_7 (y_7^*-\bar y_7^*)-x_7^*(y_7-\bar y_7)+\lambda_2'(x_4^*-x_4)~.~~~
\label{eofm}
\end{eqnarray}

We generate the non-linear parts (i. e., the parts coming from neutral current
interactions) of the 22 remaining equations (four of them redundant) from the following successive steps: 1) simultaneous  permutations of indices $123\rightarrow 456$, and $456\rightarrow 789$; 2) the same step repeated; 3) the interchange
$x\leftrightarrow y$ (bringing us up to a total of 12 equations) ; 4)
adding a bar to each unbarred symbol and removing the bar from each barred symbol, combined with an all-over change in sign of the right hand sides. The remaining linear terms on the right hand side, containing the $\nu$-mass terms and the electron interaction term are better computed by hand from (\ref{hamb}) using (\ref{com2}). 

These equations are the generalization of those derived in ref.\cite{sr}, and applied in \cite{sig}-\cite{qf}, for a situation which has only two flavors and which is spherically symmetrical.
The equations of ref \cite{sr}, specialized to two flavors and to a single energy, can be easily recaptured
from (\ref{hummer}). We can improve a little on the set (\ref{eofm}), as extended by the permutations listed above,
by remembering that the up-moving particles have, on the average, higher energies than the down-moving ones. We incorporate
this through reducing the oscillation terms for the $y$ variables by a factor $\xi=\langle E_e \rangle /  \langle E_{\mu,
\tau} \rangle $, as calculated using the \underline{initial values} of these energy averages, leading to equations of the form,
\newpage
\begin{eqnarray}
&i\dot y_1=y_1(x_9-x_6-\bar x_9+\bar x_6)-y_4^*(x_7^*-\bar x_7^*)+
\nonumber\\
&y_7^*(x_4^*-\bar x_4^*)+(y_6-y_9)(x_1-\bar x_1)+\xi \lambda_1' (y_6-y_9)+
\nonumber\\
&\xi \lambda_2' y_7^*-{n_e -\bar n_e\over 2 n_v}y_1~,
\nonumber\\
\,
\nonumber\\
&i \dot y_3=y_4^*(x_4-\bar x_4)-y_4(x_4^*-\bar x_4^*)+y_7 (x_7^*-\bar x_7^*)
\nonumber\\
&-\xi y_7^*(x_7-\bar x_7)+\xi \lambda_2' (y_4^*-y_4)~,~~{\rm etc.}
\label{eofm}
\end{eqnarray}

We take $\xi=.6$ in what follows.
Turning to the solutions, we take a neutrino density for each species, $n_v=18 ({\rm MeV})^3$ , about 1/2 the 
thermal density for a temperature of 7 MeV. Thus our picture is that roughly half of the neutrinos fall into the
``valence" group. Then the unit of
energy in our scaled equations is $E_s=2 \sqrt 2 n_\nu G_F\approx 5\times 10^{-10}{\rm MeV}$ (corresponding to a distance parameter of $\approx .04$ cm.) We take an average $\nu$ energy of $17$ MeV. The $\nu_e$ oscillation parameter, $\lambda_1'=\delta m^2_{e-\tau}/2E_\nu $, in units of $E_s$, is now $\lambda _1' \approx 6\times 10^{-9}$, for $\delta m^2_{e-\tau}=10^{-4}$ eV, a number quite unfriendly to computations since 
there are oscillations in the key flavor-changing densities at  periods of order $E_s^{-1}$. We take $\lambda_2'=50 \lambda_1'$.  For a mass density of $8 \times 10^{10}{\rm  g c^{-3}}$ and with $Y_e=.4$, we have $n_e/ n_\nu 
\approx 8$.

In fig.1 we show a solution to the equations with artificially large values of the oscillation parameters, $\lambda_1'=10^{-5}$, 
and 
$\lambda_2'$ scaled to $\lambda_1'$ as described above. We take initial conditions that correspond to six groups
of equal density, with $\nu_e$, $\bar \nu_e$ moving downward and  $\nu_{\mu,\tau}$, $\nu_{\bar \mu, \bar \tau}$
moving upward.  This initial condition translates into
initial values, $x_3=\bar x_3=x_9=\bar x_9=-1/3; x_6=\bar x_6=2/3$; and $y_3=\bar y_3=y_9=\bar y_9=1/3; y_6=\bar y_6=-2/3$, with all other initial values equal to zero.

\begin {figure}[ht]
    \begin{center}
       \epsfxsize 2.75in
        \begin{tabular}{rc}
           \vbox{\hbox{
$\displaystyle{ \, { } }$
               \hskip -0.1in \null} 
} &
            \epsfbox{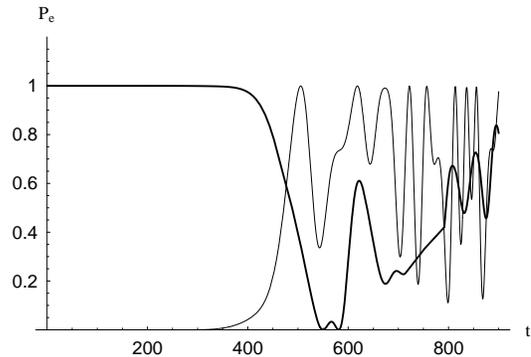} \\
            &
            \hbox{} \\
        \end{tabular}
    \end{center}
\label{fig.1}
\protect\caption
    {%
Electron neutrino density in the up-moving beam (light curve) and in the down-moving beam (heavy curve) as a
function of time, as expressed as a fraction $P_e$ of the original density $n_\nu$. The unit of time
is the fast scale, as defined in text, $t_s=(2 \sqrt 2 G_F n_\nu)^{-1}$. In this calculation the oscillation parameters have been
taken to be $1.6\times 10^3$ times as big as present oscillation parameters demand, with $\lambda_1'=10^{-5}$,
$\lambda_2'=50 \lambda_1'$, in order to better show features of the curve and
to exhibit the scaling, discussed in text, which roughly relates these curves to those of fig. 2.}
\end {figure}

From the plots of fig.1 we see that strong mixing sets in shortly after a time $t=(\lambda_1')^{-1/2}$, with the down-moving 
states at some moments nearly totally occupied with $\nu_{\mu,\tau}$. Not exactly in synch with the down-moving mixing, the up-going
states become up to 50\% occupied with $\nu_e$'s, although the percentage fluctuates wildly with time. We observe that
the total number of $\nu_e$'s and $\bar \nu_e$'s is not constant in the above, although their difference is constant. 
Owing to the perfectly
symmetrical initial condition, the plots for anti-neutrinos are identical. These results are a prototype for all that follows;
for smaller values of $\lambda_1'$, the time required for big mixing of the down-moving states follows the $t=( \lambda_1')^{-1/2}$ rule fairly closely as the oscillation parameters are reduced.

In fig 2. we show the case $\lambda_1'=10^{-7}$ (still 16 times our target value). The flipping in the up-moving states occurs after roughly $5000/(2\pi)$ periods of the
rapid oscillation rate \footnote{Luckily for our plots, this very rapid oscillation surfaces
only in the expectations of the flavor-off-diagonal density operators.} $2 \sqrt 2 n_\nu G_F$. Comparing fig.1 with fig.2, where there is a factor of 100 difference in the oscillation parameters, and a factor of 10 difference in the time spans, we see
good confirmation of the $t=(\lambda_1')^{-1/2}$ condition for mixing. Therefore it appears that we can extrapolate
reliably to the physical value of $\lambda'$. Doing so, we obtain a time of approximately $2.4\times 10^4$ in our fast units,
giving a mixing distance of around 1000 cm.

\begin {figure}[ht]
    \begin{center}
       \epsfxsize 2.75in
        \begin{tabular}{rc}
           \vbox{\hbox{
$\displaystyle{ \, { } }$
               \hskip -0.1in \null} 
} &
            \epsfbox{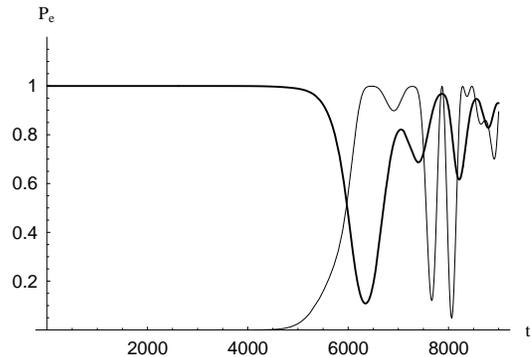} \\
            &
            \hbox{} \\
        \end{tabular}
    \end{center}
\label{fig.2}
\protect\caption
    {%
The same as fig. 1, except that each oscillation parameter is reduced by a factor
of 100, so that $\lambda_1'=10^{-7}$, etc. The unit of time is again the short time $t_s$, which is the oscillation period of the
flavor-non-diagonal operators. Comparing with fig. 1 we see the $(\lambda ')^{1/2}$ scaling discussed in text. } 
\end {figure}

In the calculation leading to fig. 3 we take a five percent $\nu_e-\bar \nu_e$ surplus in the initial state, again
for the case $\lambda_1'=10^{-7}$. 
We see from fig. 3 that in this case one-half of the up-going states become occupied with $\nu_e$, $\bar \nu_e$,
as in the previous cases, but more steadily, whereas now the occupancy of the down-moving states is changed hardly at all.
There is only a tiny difference in the behavior of the neutrino and antineutrino plots in this case, despite the fact that
major features are driven by the small initial asymmetry.

\begin {figure}[ht]
    \begin{center}
       \epsfxsize 2.75in
        \begin{tabular}{rc}
           \vbox{\hbox{
$\displaystyle{ \, { } }$
               \hskip -0.1in \null} 
} &
            \epsfbox{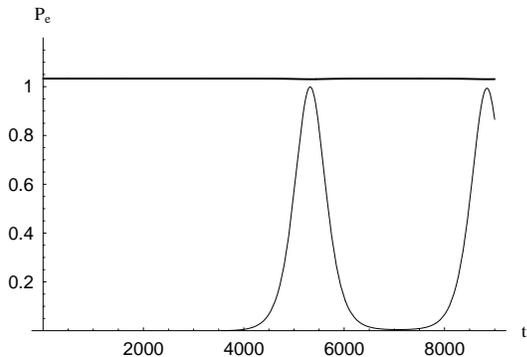} \\
            &
            \hbox{} \\
        \end{tabular}
    \end{center}
\label{fig.3}
\protect\caption
    {%
The same conditions as fig. 2, except for the addition a 5\% excess of $\nu_e$'s over the other species in the
initial up-moving beam.  }
\end {figure}

In a third situation we change the rules so that the $\bar \nu_e$'s are placed in the upward beam. Since in typical supernova
calculations the $\bar \nu_e$ are intermediate between the $\bar \nu_{\mu,\tau}$'s and the $\nu_e$'s, both in energy spectrum
and in angular distribution, this is a calculation complementary to the ones presented above. As mentioned earlier, it corresponds to a different choice for the ``core" states. Again using values of
$\lambda_1'$ and $\lambda_2'$ that  are 16 times too large, because of computational constraints, we show in fig. 4 results
for the electron neutrino distributions, in this case appearing to be a clean flip. In contrast to the earlier cases, shown in figs. 1-3,  the anti-neutrino distributions are frozen at their initial values in this case.
  
\begin {figure}[ht]
    \begin{center}
       \epsfxsize 2.75in
        \begin{tabular}{rc}
           \vbox{\hbox{
$\displaystyle{ \, { } }$
               \hskip -0.1in \null} 
} &
            \epsfbox{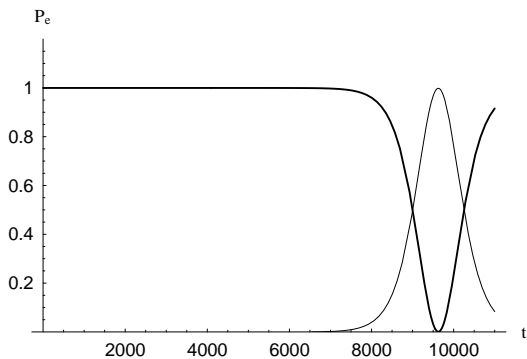} \\
            &
            \hbox{} \\
        \end{tabular}
    \end{center}
\label{fig.4}
\protect\caption
    {%
The same parameters as for fig. 2, except that in the initial conditions the $\bar \nu_e$'s
have been put in the up-moving group.
 }
\end {figure}

\section{Discussion}
The purposes of this paper were:

 1) To give a simple formulation of the complete 3-flavor equations
that govern neutrino evolution, over time periods that are short compared
to scattering times, when the density of neutrinos and antineutrinos is large. This is accomplished in
the construction of the forward neutral-current Hamiltonian (\ref{ham}) and the commutation rules 
(\ref{commutators}), used in conjunction with the usual oscillation terms. Of course, in a complete approach
with radically non-monoenergetic distributions, the densities have still another index, $E$, and we would write equations for
densities $\rho^{(E)} (\Omega)$. In this case the forward neutral current Hamiltonian has a double sum over energies but no explicit 
further energy dependence. The energy enters explicitly in the oscillation terms. 
 
2) To find a way of treating these equations in the quasi-monoenergetic case,
a case that at least can give some insight into possible behaviors. It appears to us that the approximation of
using a few rays in the space of angles of the momentum distribution, each with its own complete set of
20 densities is both conceptually clear and practical to implement, the only caveat being that going beyond 
the two-ray approximation of the present paper will require large computing resources, owing to the
extreme values of dimensionless parameters. A possible alternative
would be to make a partial wave expansion of the density operators. Of course, because of the non-linearity,
and the angular factor $1-\cos \theta _{\Omega, \Omega '}$ in the kernel of (\ref{ham}), multipoles of all
orders are coupled together. Thus, even though all reactions are forward, so that the manifold of occupied momentum states
remains unchanged, and even though we might choose an initial configuration with, say, only $l=0,1,2$ 
states present in the distributions, the \underline{flavor} configuration could, over time, become
arbitrarily complex in angle, developing modes of higher and higher multipolarity. In any case, because
we cannot verify the stability of the truncation procedure we do not pursue the partial wave approach here, preferring
instead the discrete rays approach. In the rays approach, we represent continuous distributions in the initial state less well, but 
we really solve the resulting dynamical evolution problem. On the basis
of the previously investigated two-flavor models we expect that when we begin with more than two rays
 there will be instabilities
that grow on a time scale of several times $(G_F n_\nu) ^{-1}$ rather than $(G_F n_\nu \lambda_1)^{-1/2}$, under the conditions
of this present paper, perhaps rendering all of our present results academic \footnote{
This is also true with a truncated partial wave expansion that includes $l=0,1$ and $2$ states.}. Unfortunately, we are not 
equipped to deal computationally with the generalization of the SU3 case, with antineutrinos included, to a case 
with more than the two rays of the present paper.

Turning to the specific results of the present paper: In three different idealizations, each capturing some of
the features that could be important in the supernova neutrinosphere problem, we find an
array of possible behaviors. In each case the time-scale is of the order $ (E_s \lambda_1)^{-1/2}$, corresponding
to a reaction distance of order 100 cm. In the first case we took an initial condition in which both $\nu_e$ and
an equal number of $\bar \nu_e$ from the valence group were assigned to the down-going beam, while 
 $\nu_\tau$,
$\bar\nu _\tau$, $\nu_\mu$, $\bar\nu _\mu$ were assigned to the up-going beam in equal numbers. 
The result was a nearly complete flavor over-turn, in which at some moments in time one-half of the up-moving
beam is composed of $\nu_e$'s and the down-moving beam is entirely composed of $\nu_{\mu,\tau}$'s.

In the next example, we added a small extra number of $\nu_e$'s in the initial
down-moving group, uncompensated by anti-particles. In this case the system does a very amusing thing.
The down-moving states remain occupied by $\nu_e$, $\bar \nu_e $ ; but the up-moving states nevertheless make
excursions into a configuration in which half of the total original number of $\nu_\mu$'s, $\nu_\tau$'s
have transformed into $\nu_e$'s, and similarly for the antiparticles.

In the final example we put the initial $\bar \nu_e$'s into the up-moving group. In this case
there is perfect flavor neutrality in the anti-neutrino sector (i.e. a nonvanishing expectation
only for the operator $\bar y_0$), and no antineutrino action. However, there is big mixing of the neutrinos in the (moderately) fast time scale, $(E_s \lambda_1)^{-1/2}$,
as we saw in fig.4. 

One thing in common in all of these simulations is that they require \underline{both} parameters $\lambda_1$ and $\lambda_2$
to be non-zero, even though when $\lambda_1<<\lambda_2<< G_F  n_\nu$ the effective time constant, $ (E_s \lambda_1)^{-1/2}$,
does not depend greatly on $\lambda_2$.
Thus all of the phenomena manifesting themselves here \underline{require} having three flavors and two mixings.

We can try to separate the effects of the two innovations in this paper: 1)
the two-group treatment of the angular effects; 2) the inclusion of all three flavors and their antiparticles. We perform
all of the calculations described above but this time beginning from the completely isotropic equations. There are
nearly  twice as many terms on the right hand sides of the equations, as compared to (\ref{eofm}), owing to the
intra-group interactions \footnote{A group now consists of a set of neutrinos with momenta in all directions. In the
neutral current interaction $\cos \theta$ is replaced by its average value of zero.
We still use two groups of states, to take into account the different energy spectra.}
In the analogs of the calculations with completely symmetric conditions that led to figs 1,2 , we find much diminished
mixing effects at approximately ten times the time as in the case of our basic calculation. For the other two
examples, which led to the plots of fig. 3 and fig. 4., the isotropic equations gave no appreciable mixing. 

It would be rash to extrapolate from any of the above results to draw firm conclusions for supernovas. Since our exploratory 
work \cite{rfs1} in the two-flavor case indicates the existence of instabilities that can lead to macroscopic changes 
on a \underline{shorter} time scale than $(E_s \lambda_1)^{-1/2}$, once we introduce more complex angular dependence, 
someone with massive computing power should explore this case. The results could render much of the above obsolete, but not the basic equations
or the method of attack. A reader could ask, ``Why spend so much effort on a problem, if the author believes that it
misses the main points of the ultimate solution?" The answer is that this appears to be quite a hard problem, and that
systematic work around the edges may be likelier to bring us an understanding of what needs to be done than is plunging
ahead, discarding this and that and approximating the other, in order to claim a conclusion for a physical system.  

The main moral of our present results, as well as those presented in ref. \cite{rfs1}, is that there are big nonlinear effects
latent in the neutral-current couplings, effects
which can lead to what we have called speed-ups. We speculate that there is a connection
between the speed-ups of the present paper, which we would say are due to marginal instabilities, and those more rapidly
growing instabilities described in ref. \cite{rfs1}. We think that the examples considered above sit exactly on the boundary
of an unstable region in a parameter space. Indeed, the route to finding the growing modes in ref. \cite{rfs1} was
to take seriously a demonstration that the two-beam case (as defined in the present paper) is on such a boundary. 
If the author were to guess at the ultimate outcome, he would guess that
the end result of the non-linear effects will be total flavor mixing, and on a distance scale \underline{shorter}
than the $1000$cm. that the calculations of this paper indicate. This would be perhaps a rather dull outcome, but
it would be computationally convenient as far as incorporating the answers into big simulations of the complete supernova process are concerned.

We emphasize that there is no MSW transformation in any of our results. We deal in time
spans that are so short that the density of the surrounding medium can be taken as constant, in the first place.
In the second place, the effective time for the actual turn-over, after the waiting time $(E_s \lambda_1)^{-1/2}$,
during which the system prepares itself, is many orders of magnitude faster than an MSW transition time
can be for any time-changing (i.e. space-changing) medium, for the oscillation parameters that we choose. This distinguishes
our considerations from those of refs. \cite{sig}-\cite{qf}.

\section{Appendix}

The coefficients $f_{i,j,k}$ which in our representation are antisymmetric only in the first two indices,
have the following non-vanishing components, 

\begin{eqnarray}
&f_{1,2,6}=f_{1,4,8}=f_{1,9,1}=f_{2,6,2}=f_{2,8,4}=f_{3,5,5}=
\nonumber\\
&f_{3,7,7}=f_{4,5,9}=f_{4,7,2}=f_{5,9,5}=
f_{6,8,8}=f_{7,8,3}=1 \, ,
\nonumber\\
\,
\nonumber\\
&f_{1,2,9}=f_{1,6,1}=f_{1,7,5}=f_{2,5,7}=f_{2,9,2}=f_{3,4,4}=
\nonumber\\
&f_{3,8,8}=f_{4,5,3}=f_{4,9,4}=f_{5,8,1}=
f_{6,7,7}=f_{7,8,6}=-1\, ,
\nonumber\\
\,
\end{eqnarray}
as well as the components obtained by interchange of the first two indices. Although this set is slightly less economical
than the canonical set of Gell-Mann (which are based on a hermitean set of $\rho_i$),
 the present representation serves us well for the purpose of writing down the Heisenberg equations.
In this representation it suffices to write only the equations for the time derivatives of $\rho_1$ and $\rho_3$,
with the rest found by permuting indices, as we discussed earlier. The non-hermitean operators
that we use also made it easy for us to express the quantities $\langle \rho_2 \rangle,\, \langle \rho_5\rangle ,\, \langle
\rho_8 \rangle $ in terms of
the respective complex conjugates of $\langle \rho_1 \rangle ,\, \langle \rho_4\rangle ,\, \langle \rho_7 \rangle$, reducing the number of integro-differential equations from sixteen to ten, or the number of independent ordinary DE's from thirty two to twenty, in the case of our two-beam approach.  

We show the first five of these equations, beginning with the two presented previously. The remaining 15, now including
the mass terms, are obtained from the final two substitution rules given in text: a.) the interchange
$x\leftrightarrow y$; b.)
adding a bar to each unbarred symbol and removing the bar from each barred symbol, combined with an all-over change in sign of the right hand sides, 
\begin{eqnarray}
&i\dot x_1=x_1(y_9-y_6-\bar y_9+\bar y_6)-x_4^*(y_7^*-\bar y_7^*)+
\nonumber\\
&x_7^*(y_4^*-\bar y_4^*)+(x_6-x_9)(y_1-\bar y_1)+\lambda_1' (x_6-x_9)+
\nonumber\\
&\lambda_2' x_7^* -{n_e -\bar n_e\over 2 n_v}x_1,
\nonumber\\
\,
\nonumber\\
&i \dot x_3=x_4^*(y_4-\bar y_4)-x_4(y_4^*-\bar y_4^*)+x_7 (y_7^*-\bar y_7^*)~~~~~~~~~~~~~~~~
\nonumber\\
&-x_7^*(y_7-\bar y_7)+\lambda_2'(x_4^*-x_4)~,~~~~~~~~~~~~~~~
\nonumber\\
\,
\nonumber\\
&i\dot x_4=x_4(y_3-y_9-\bar y_3+\bar y_9)-x_7^*(y_1^*-\bar y_1^*)+
\nonumber\\
&x_1^*(y_7^*-\bar y_7^*)+(x_9-x_3)(y_4-\bar y_4)+
\nonumber\\
&-\lambda_1' x_7^*+\lambda_2' (x_9-x_3) \, ,
\nonumber\\
\,
\nonumber\\
&i \dot x_6=x_7^*(y_7-\bar y_7)-x_7(y_7^*-\bar y_7^*)+x_1 (y_1^*-\bar y_1^*)~~~~~~~~~~~~~~~~
\nonumber\\
&-x_1^*(y_1-\bar y_1)+\lambda_1'(x_1-x_1^*)~,~~~~~~~~~~~~~~~
\nonumber\\
\,
\nonumber\\
&i\dot x_7=x_7(y_6-y_3-\bar y_6+\bar y_3)-x_1^*(y_4^*-\bar y_4^*)+
\nonumber\\
&x_4^*(y_1^*-\bar y_1^*)+(x_3-x_6)(y_7-\bar y_7 )
\nonumber\\
&\lambda_1' x_4^*-\lambda_2' x_1^* + {n_e -\bar n_e \over 2 n_v}x_7\, .
\label{eofm2}
\end{eqnarray}

\end{document}